\documentclass[english,aps,preprint]{revtex4}

\usepackage{babel}
\usepackage{amsmath}
\usepackage{amsfonts}
\usepackage{amssymb}

\def\CC{\mathbb C} \def\RR{\mathbb R} \def\ZZ{\mathbb Z} \def\NN{\mathbb N} 
 
\def\id{ 1\kern-4pt 1}
 
\def\L{{\mathcal L}}
\def\H{{\mathcal H}}
\def\K{\mathcal K}
 \def\M{\mathcal M} \def\A{\mathcal A}   
  \def\C{\mathcal C}  \def\F{\mathcal F}  
\def\uz{\underline{z}}
\def\uzo{{\underline{z}}_{0}}
\def\Hcl{H_{cl}}
\def\Hocl{H_{0,cl}}
\def\pp(#1,#2){\frac{\partial #1}{\partial #2}}
\def\Gen{G}\def\AA{\Lambda}
\def\hbara{\hbar}
\def\hbarb{\hbar}
\def\alh{\lambda}
\newcommand\nl{\hfill\break}

\begin{document}

\title{Dynamics of mixed classical-quantum systems, 
geometric quantization and coherent states \footnote{published in ``Mathematical Horizons for Quantum Physics'', 
Huzihiro Araki, Berthold-Georg Englert, Leong-Chuan Kwek, Jun Suzuki (Eds.); 
IMS-NUS Lecture Notes Series, Vol 20, World Scientific, Singapore, 2010}}
\author{H.R. Jauslin, D. Sugny}
\address{Institut Carnot de Bourgogne UMR 5209 CNRS,
Universit\'e de Bourgogne, BP 47870, 21078 Dijon, France}
\date{\today} 

\begin{abstract}
We describe quantum and classical Hamiltonian dynamics in a common Hilbert space framework,
that allows the treatment of mixed quantum-classical systems. The analysis of some examples illustrates the possibility of entanglement between classical and quantum systems. We give a summary of the main tools of Berezin-Toeplitz and geometric quantization, that provide a relation between the classical and the quantum models, based essentially on the selection of a subspace of the classical Hilbert space.
Coherent states provide a systematic tool for the inverse process, called dequantization, that associates a  classical Hamiltonian system to a given quantum dynamics through the choice of a complete set of coherent states.
\end{abstract}

\maketitle

\tableofcontents
\newpage

\section{Introduction}

In order to describe the dynamics of a bipartite system, in which 
one part is quantum and the other one is classical, we have to describe
the classical and quantum systems in a common mathematical framework.
The construction is guided by the idea that we can consider the classical
system as a limit, and as a particular case, of a quantum system. The
framework we consider is provided by the Koopman-von Neumann 
\cite{Koopman-1931,vonNeumann-1932} representation
of classical mechanics combined with a quantization procedure of the
Berezin-Toeplitz type. 

This construction is to be distinguished from models in which the coupling between a classical system (e.g. a classical electromagnetic field)  and a quantum system composed of many identical atoms is described by a coupling involving the average of some quantum observable (e.g. the electric dipole moment). This type of effective models is useful e.g.  to describe the induced classical field produced  by  a large number of atoms excited by the initial electromagnetic wave. It is implicitly assumed that the  quantum average provides an effective description of the collective retroaction of the atoms on the classical system. The models we discuss in this article are meant to describe the dynamics of a single quantum system in interaction with a single classical object. A prototypical example is a model for the Stern-Gerlach experiment in which the motion of the center of mass of the atom is treated classically and the spin is treated quantum mechanically.

In Section \ref{HQM-sect} we summarize the mathematical framework of quantum dynamics. In Section 
\ref{Hcl-sect} we describe a formulation of classical mechanics in a Hilbert space framework, on the basis of the formulation of Koopman and von Neumann. Once classical and quantum dynamics are formulated in the same mathematical framework, in Section \ref{mixed-sect} we describe the formalism to treat mixed systems, in which one part is classical and another one is quantum. We illustrate this formalism with a simplified model for the Stern-Gerlach experiments, in which the motion of the center of mass is considered as classical and the spin quantum mechanical. Section \ref{geometricQ-sect} presents with a minimum of mathematical formalism some of the main ideas of geometric and Berezin-Toeplitz quantization. This allows to establish a well-defined relation between classical and quantum models. The main idea is that one can obtain quantum models from the Koopman - von Neumann representation of classical dynamics in a Hilbert space just by selecting a suitable subspace of the classical Hilbert space. In Section \ref{dequantization-section} we describe the inverse process, called dequantization, in which starting with a given quantum system and a set of coherent states one constructs a classical phase space and Hilbert space, a corresponding polarization subspace, classical observables and a classical dynamics.

\section{\label{HQM-sect}Hilbert space framework of quantum dynamics}

A quantum dynamical system is defined by the following elements:\nl
(i) A separable Hilbert space $\H$. \\
(ii) An algebra of observables $\A$ and a representation $\rho(\A)$ as linear operators on the Hilbert space $\H$. The physical observables are given by the self-adjoint elements.\\
(iii) The dynamics, in the Heisenberg representation, is defined by a derivation operator $D_{\hat H}$, that acts on the algebra  $\rho(\A)$. The derivation is constructed from the Hamiltonian $\hat H$, which is a particular self-adjoint element of the algebra $\A$.
The derivation $D_{\hat H}$ can be expressed in terms of the commutator with a linear operator $\hat H$ acting on the Hilbert space $\H$:
$D_{\hat H}=i[\frac{1}{\hbar}\hat H,~.~]$. The Heisenberg equation for an observable $\hat A$ can be written as
$$
\frac{\partial \hat A}{\partial t} = D_{\hat H} (\hat A)=i[\frac{1}{\hbar}\hat H, \hat A].
$$
(iv) The Hamiltonian $\hat H$  defines also the dynamics of the states $\psi\in\H$ in the 
Schr\"odinger picture, by 
$$
i\frac{\partial \psi}{ \partial t}= \frac{1}{\hbar}\hat H \psi.
$$
We remark that, in order to prepare a common notation with the classical systems, all the dependence on Planck's constant $\hbar$ is attached to the operators.

\section{\label{Hcl-sect}Hilbert space framework of classical dynamics}

The Koopman-von Neumann \cite{Koopman-1931,vonNeumann-1932,Deotto-Gozzi-Mauro-JMP2003} formalism is a representation
of classical dynamics on a phase space $\M$ in terms of a Hilbert
space of square integrable functions $\L_{K}:=L_2(\M, d\mu)$, with respect to a measure $d\mu$. The classical
observables, which are differentiable functions $f:\M\to \CC$ are represented on $\L_K$ by
a commutative algebra of multiplication operators $M_f$: For $\xi\in\L_K$
$$
(M_f\xi)(\uz):=f(\uz)\xi(\uz).
$$
In order to simplify the notation we will sometimes write  $f$ instead of $M_f$.

\subsection{Koopman-Schr\"odinger representation}

The classical time evolution is determined by a Hamilton function
$\Hcl(\uz)$, where we denote the local coordinates by $\uz=(p,q)\in\M$, $p=p_1,\ldots,p_d$, $q=q_1,\ldots,q_d$.  
The classical Hamiltonian flow, denoted by $\uz(t)=\Phi_t(\uz_0)$,  which satisfies the classical Hamilton equations
\begin{equation} \label{Hamiltoneqs}
 \frac{dp}{ dt} =-\pp(\Hcl,q),\qquad \frac{dq}{ dt} =\pp(\Hcl,p), 
\end{equation}
which can be written as
$$
\frac{d\Phi_t(\uz_0)}{ dt} =J~\nabla \Hcl (\Phi_t(\uz_0)); \qquad J=\left(\begin{array}{cc}0 & -1 \\1 & 0\end{array}\right),  \qquad \nabla=\left(\begin{array}{c}\frac{\partial}{ \partial p} \\ \frac{\partial}{ \partial q}\end{array}\right)
$$
can be used to define a unitary flow on the Koopman Hilbert space
\cite{Koopman-1931,vonNeumann-1932,Reed-Simon-1}: 
$ U^K(t,t_0)\xi_0(\uz):=\xi_0(\Phi_{-(t-t_0)}(\uz))$. We  will use a slightly different  unitary flow, which we will call Koopman-Schr\"odinger flow, that includes a time-dependent phase: 
\begin{equation}
\label{koopman-op}
\xi(\uz,t):=U^{KS}(t,t_0)\xi_0(\uz):=e^{-i\delta(\uz,t-t_0)}\xi_0(\Phi_{-(t-t_0)}(\uz)),\qquad \xi(\uz,0)=\xi_0(\uz),
\end{equation}
with
\begin{equation}
\label{koopman-phase}
\delta(\uz,t-t_0)=\frac{1}{\hbara}\Biggl((t-t_0)~\Hcl(\uz)+\int_{t_0}^t dt'~\AA_{\Hcl}(\Phi_{t'-(t-t_0)}(\uz))\Biggr),
\end{equation}
where the function $\AA_{\Hcl}(p,q)$ will be chosen as \cite{Souriau-geometric-quantization-1970,Kostant-geometric-quantization-1970,Woodhouse-book-1992,Faure-JMD-2007} 
\begin{equation}
\label{gauge_potential}
\AA_{\Hcl}(p,q)=-\frac{1}{2}\sum_j\Biggl(p_j\pp(\Hcl,p_j)+q_j\pp(\Hcl,q_j) \Biggr).
\end{equation}
The flow  (\ref{koopman-op})  with the phase (\ref{koopman-phase}) is called the {\it prequantum flow}  in the framework of {\it geometric quantization~}\cite{Souriau-geometric-quantization-1970,Kostant-geometric-quantization-1970,Woodhouse-book-1992}.
The first term in (\ref{koopman-phase}) is called a {\it dynamical phase} and the second one a {\it geometrical phase}.
The constant $\hbara$ introduced in (\ref{koopman-phase}), which has the dimension of an action ([time]$\cdot$[energy]) is necessary in order to have dimensionless numbers in the exponential. Its numerical value has to be determined by comparison between the predictions of the quantum or of the mixed models with the corresponding experimental measurements.
\\ \\
The interpretation of the classical wave function $\xi(p,q,t)$ is that  $\rho d\mu:=|\xi(p,q,t)|^2d\mu$ gives the probability density at time $t$ for a particle to have a momentum $p$ and a position $q$. This interpretation comes from the fact that $\rho$ is a real positive function that satisfies the Liouville equation 
$\pp(\rho,t)=-\{\Hcl,\rho\}$, which is the equation that describes the time evolution of the probability densities $\rho$ in phase space defined by the flow of Eqs. (\ref{Hamiltoneqs}). Indeed Eq (\ref{koopman-op}) implies that  
$ \rho(\uz,t)=\rho_0(\Phi_{-(t-t_0)}(\uz))$, with $\rho_0:=|\xi_0|^2$, and thus $\pp(\rho,t)=-\{\Hcl,\rho\}$, where the brackets $\bigl\{~,~.\bigr\}$ denote the Poisson brackets
defined by
$
\bigl\{h,f\bigr\}:=\sum_j \frac{\partial h}{\partial p_j}\frac{\partial f}{\partial q_j}-\frac{\partial h}{\partial q_j}\frac{\partial f}{\partial p_j}.
$\\\\
For a purely classical system this global phase $\delta(\uz,t)$ is thus irrelevant since it is 
the square of the absolute value  $|\xi(\uz,t)|^2$ that gives all the physically relevant quantities. We will see that the analogue of the phase term plays an essential role in the description of  quantum systems and of mixed classical-quantum systems.
\\ \\
By Stone's theorem \cite{Reed-Simon-1}, this unitary flow has a selfadjoint generator
$\Gen_{\Hcl}$, i.e. it can be written as
\begin{equation}
U^{KS}(t,t_0)=U^{KS}(t-t_0)=e^{-i(t-t_0) \Gen_{\Hcl}}.  
\end{equation}

The dynamics of the vectors of the Hilbert space satisfies therefore
the following Schr\"odinger type equation, which we will call the Koopman-Schr\"odinger
equation:
\begin{equation}
\label{Koopman-Schroedinger-eq}
 i\frac{\partial\xi}{\partial t}=\Gen_{\Hcl}\xi. 
\end{equation}
The generator $\Gen_{\Hcl}$ can be written explicitly as the sum of three operators
\begin{equation}
\label{Generatora}
\Gen_{\Hcl} = \tilde M_{\Hcl}+\tilde M_{\AA_{\Hcl}}+\tilde X_{\Hcl},
\end{equation}
or explicitly,
\begin{equation}
\label{Generatorb}
\Gen_{\Hcl} = \frac{1}{\hbara} \Hcl-\frac{1}{2\hbara}\sum_j \Bigl(p_j\pp(\Hcl,{p_j}) +
q_j\pp(\Hcl,{q_j}) \Bigr) -i \sum_j \frac{\partial \Hcl}{\partial p_j}\frac{\partial}{\partial q_j}-\frac{\partial \Hcl}{\partial q_j}\frac{\partial}{\partial p_j},   
\end{equation}
where $\tilde M_{\Hcl}:=\frac{1}{\hbar}M_{\Hcl}$ is the multiplication operator with the Hamilton function, $\tilde M_{\AA_{\Hcl}}:=\frac{1}{\hbara}M_{\AA_{\Hcl}}$ is the multiplication operator with $\frac{1}{\hbara}\AA_{\Hcl}(p,q)$,  and 
$\tilde X_{\Hcl}:=-iX_{\Hcl}$ is the vector field associated to the Hamilton function ${\Hcl}$, i.e. the differential operator 
$$
\tilde X_{\Hcl}:=-i X_{\Hcl}=-i \sum_j \frac{\partial \Hcl}{\partial p_j}\frac{\partial}{\partial q_j}-\frac{\partial \Hcl}{\partial q_j}\frac{\partial}{\partial p_j} \equiv -i \bigl\{\Hcl,~.\bigr\}.
$$

The dynamics of the classical system defined by Eq. (\ref{Koopman-Schroedinger-eq}) is 
called {\it prequantum dynamics}, since, as we we will discuss in Section 
\ref{quantization-of-dynamics}, it is the starting point for the construction of the quantum dynamics in the framework of geometric quantization 
\cite{Souriau-geometric-quantization-1970,Kostant-geometric-quantization-1970,Woodhouse-book-1992}.
The motivation for the addition of the terms $ \tilde M_{\Hcl}$ and $\tilde M_{\AA_{\Hcl}}$ in (\ref{Generatorb}) is:\\
(i) The term   $ \tilde M_{\Hcl}$ is added in order to have $\Gen_{f(\uz)=1} =\frac{1}{\hbara} \id$. If we had only 
$\Gen_{\Hcl} = \tilde X_{\Hcl}$ it would lead to  $\Gen_{f(\uz)=1} = 0$, which would not allow a systematic construction of the dynamics of mixed systems.\\
(ii) $\tilde X_{\Hcl}$ satisfies $ \tilde X_{\{f,g\}}= i[\tilde X_f,\tilde X_g]$, but this property is not satisfied by 
$\tilde M_{\Hcl}+\tilde X_{\Hcl}$.
The motivation of the choice (\ref{gauge_potential}) for $\AA_{\Hcl}$ is that the operator
$
\nabla_{\Hcl}:= M_{\AA_{\Hcl}}+X_{\Hcl}
$
can be interpreted as a covariant derivation associated to $\Hcl$ 
\cite{Woodhouse-book-1992,Faure-JMD-2007}, and as a consequence $\Gen_g$ satisfies
$$
\Gen_{\{f,g\}}= i[\Gen_f,\Gen_g].
$$
{\bf Remarks:}  
1) The action of the Poisson brackets is defined both
on the algebra of observables (functions $f\in\A$, and their representation
as multiplication operators $\rho(f)=M_f$ ) and on the (differentiable) vectors
of the Koopman Hilbert space $\L_K$. 
\\
2) The operators $ M_{\Hcl}$ and $X_{\Hcl}$ commute, since
$$
 \bigl[M_{\Hcl}~,~X_{\Hcl}\bigr]\xi = \Hcl \bigl\{\Hcl,\xi\bigr\} - 
 \bigl\{\Hcl,\Hcl \xi\bigr\}
$$
and $ \bigl\{\Hcl,\Hcl \xi\bigr\}=\Hcl \bigl\{\Hcl, \xi\bigr\}+ \bigl\{\Hcl,\Hcl \bigr\}\xi=\Hcl \bigl\{\Hcl, \xi\bigr\}$.\\
\\
However, except for special choices of the Hamiltonian $\Hcl$, the operators $ M_{\AA_{\Hcl}}$ and $X_{\Hcl}$ do not  commute.

The unitary  Koopman-Schr\"odinger operator (\ref{koopman-op}),(\ref{koopman-phase}) can be written as
$$
U^{KS}(t)=e^{-i\delta} e^{-i(t-t_0) X_{\Hcl}}=e^{-i(t-t_0) \Gen_{\Hcl}}.
$$
Some examples of operators $\Gen_{\Hcl}$ are given in Table \ref{table-G}.

\subsection{Koopman-Heisenberg representation}

One can also define a dynamics on the algebra of observables, which we will call the Koopman-Heisenberg representation,  by 
$$
M_f(t):= (U^K(t))^{-1} M_f  U^K(t)=e^{it \Gen_{\Hcl}}  M_f  e^{-it \Gen_{\Hcl}}.
$$
Since $U^K(t)=e^{-i\delta} e^{-it \tilde X_{\Hcl}}$, it can be also written as
$$
M_f(t)= e^{it\tilde X_{\Hcl}} M_f  e^{-it\tilde X_{\Hcl}}.
$$
It satisfies the following Koopman-Heisenberg equation
\begin{equation}
\label{Koopman-Heisenberg-1}
\frac{\partial M_f}{\partial t}= i [ \Gen_{\Hcl},M_f],
\end{equation}
which can be written equivalently as
\begin{equation}
\label{Koopman-Heisenberg-2}
\frac{\partial M_f}{\partial t}= i [ \tilde X_{\Hcl},M_f] = M_{\{\Hcl,f\}}.
\end{equation}
The last equality can be understood by letting it act  on an arbitrary vector $\xi\in\L_K$:
\begin{eqnarray*}
 i [ \tilde X_{\Hcl},M_f] \xi &=& i \tilde X_{\Hcl} (M_f\xi) -\tilde X_f (i H_{\Hcl}(\xi))\\ &=&
\{{\Hcl}, (M_f \xi)\} -M_f (\{{\Hcl}, \xi\}) = \{{\Hcl}, (f \xi)\} -f \{{\Hcl}, \xi\}\\
&=&\{{\Hcl}, f\}  \xi  +f\{{\Hcl},\xi\}   -f \{{\Hcl}, \xi\} \\&=& \{{\Hcl}, f\}  \xi =  M_{\{{\Hcl}, f\}}  \xi.
\end{eqnarray*}
We remark that the Koopman-Heisenberg dynamics does not involve the phase $\delta(\uz,t)$ we introduced in (\ref{koopman-op}). The Koopman-Heisenberg equation (\ref {Koopman-Heisenberg-2}) is equivalent to the equation for the time evolution of a classical observable $f(p,q)$
$$
\frac{df}{dt}=\{\Hcl,f\},
$$
but written in terms of the corresponding multiplication operator. Thus, the Koopman-Heisenberg equations for the particular observables $p$ and $q$ are equivalent to Hamilton's equations \\
We remark that the generator of the dynamics is an operator that does not belong to the algebra of observables, but it is an external derivation acting on them.  Indeed for a  Hamilton function ${\Hcl}$ the dynamics is defined by the derivation $D_{\Hcl}$, acting on the algebra $\rho(\A)$ by 
$$
D_{\Hcl} (M_f):= i [ X_{\Hcl},M_f] \equiv M_{\bigl\{{\Hcl},f\bigr\}}.
$$
This  provides a complete framework in which classical mechanics can be considered, from the mathematical point of view, as a particular type of  quantum system.\\
This formulation opens up the possibility to construct models of a quantum system in interaction with a classical one.

\section{\label{mixed-sect} Dynamics of mixed classical-quantum systems }

We consider a bipartite system composed of a quantum system (Q), defined on a Hilbert space $\H_Q$ and a classical system (K) with a Koopman Hilbert space $\L_K$.
The total Hilbert space will be $\K=\L_K\otimes \H_Q $.  The observables of the total system will be linear combinations of operators of the form $M_f\otimes \hat B$, where $\hat B$ is a selfadjoint operator of the quantum system.
We first consider the two subsystems without coupling. 
The evolution of the quantum subsystem is defined by a Hamiltonian $\id_{K}\otimes \hat H_{0}^Q/\hbar $ and the one of the classical subsystem by $ {\Gen_{\Hocl}}\otimes \id_Q/\hbar$, 
with 
$\Gen_{\Hocl}= \tilde M_{{\Hocl}}+\tilde{\AA}_{{\Hocl}}+ \tilde X_{{\Hocl}}$, and ${\Hocl}$ is the uncoupled classical Hamilton function.
\\ \\
The interaction between the classical and the quantum subsytems is determined by observables that are linear combinations of terms of the form
$$
 M_{g_i}\otimes \hat\mu_i/\hbar
$$
where $\hat\mu_i$ is a self-adjoint operator defined on $\H_Q$ and  $g_i=g_i(p,q)$ is a function on the classical phase space.

The corresponding dynamics of the interaction is defined by operators of the form
$$
K_{int}=  \sum_i \Gen_{g_i}\otimes \hat\mu_i/\hbar
$$
with $\Gen_{g_i}= \tilde M_{g_i}+\tilde{\AA}_{g_i}+\tilde X_{g_i}$,
 i.e. $ \Gen_{g_i}\xi= \frac{1}{\hbara}g_i\xi -\frac{1}{2\hbara}\Biggl(p\pp(g_i,p)+q\pp(g_i,q) \Biggr)\xi
 -i\bigl\{g_i,\xi\bigr\}$.
 
 The Schr\"odinger-Koopman equation for a mixed classical-quantum system can thus be written as
 $$
 i\pp( \psi,t) = K\psi,
 $$
with
$$
K=\id_{cl}\otimes \hat H_{0,Q}/\hbar + \Gen_{\Hocl}\otimes\id_Q/\hbar+\sum_i\Gen_{g_{i}}\otimes\hat\mu_i/\hbar.
$$
The corresponding Heisenberg-Koopman equation for the dynamics of an observable of the form
$$
B(t):= e^{itK}\bigl(M_f\otimes \hat A \bigr)e^{-itK}
$$
is
$$
\pp(B,t)=i[K,B].
$$

\subsection{Example: Stern-Gerlach experiment}

We consider a simple model of the Stern-Gerlach experiment, in which the motion of the center of mass of the atom is described classically, and the spin as a quantum variable.
The total Hilbert space is $\K=  \L_2(\RR^6, d^3p~d^3q) \otimes \CC^2$.
The states $\psi\in\K$  can be represented by 
$$ 
\psi=v_+(\vec p,\vec q)\otimes |+\rangle + v_-(\vec p,\vec q)\otimes |-\rangle 
\equiv \dbinom{v_+(\vec p,\vec q)}{v_-(\vec p,\vec q)}.
$$
The observable corresponding to the total energy is
$$
H=\frac{1}{ 2 m}{\vec p}^{\ 2}\otimes \id -\gamma \vec B(\vec q) \otimes \vec S,
$$
where $S_i=(\hbar/2) \sigma_i$ and $\sigma_i$ are the Pauli matrices.\\
The  operator that generates the corresponding  dynamics is
$$
 K =\frac {1}{ 2 m}\sum_i \Gen_{p_i^2} \otimes \id/\hbar -\gamma\sum _i \Gen_{B_i}\otimes  S_i/\hbar.
$$

As a simple model we take a magnetic field of the form  $\vec B= (0,0,B_3)$ with $B_3(\vec q)=\tilde b_0-\tilde b_1 q_3$, and $\vec q=(q_1,q_2,q_3)$. We remark that this $\vec B$ cannot be an actual  magnetic field, since $\nabla\cdot \vec B=-b_1\not=0$. We use it only as the simplest mathematical illustration of the types of behavior that can be expected. For a recent discussion of realistic models of the Stern-Gerlach experiments see Refs. \cite{Gallup-Stern-Gerlach-2001,Potel-Barranco-Stern-Gerlach-QM-2005,Cruz-Barrios-Stern-Gerlach-scl-2000}.

Since the gradient of the magnetic field is only in the $q_3$ direction, we can restrict the model to one dimension. The Hilbert space is  $\K=  \L_2(\RR^2, dp~dq) \otimes \CC^2$, where we denote $p\equiv p_3$,  $q\equiv q_3$. Absorbing $-\gamma$  into the coefficients,
the observable corresponding to the total energy can be written as
$$
H=\frac{1}{ 2 m} p^2\otimes \id +(b_0+b_1q) \otimes \sigma_3.
$$
The corresponding Koopman-Schr\"odinger equation is 
$$
i\pp(\psi,t)=\Bigl(\frac{1}{ 2 m} \Gen_{p^2}\otimes \id +(b_0\Gen_{1}+b_1\Gen_{q}) \otimes \sigma_3\Bigr)\psi.
$$
If we write $\psi(t)=({v_+(p,q,t)},{v_-(p,q,t)})$  the above equations become two independent linear partial differential equations of first order:
\begin{eqnarray}
\pp(v_+,t)& = & -\frac{p}{ m}\pp(v_+,q)+b_1\pp(v_+,p)-\frac{i}{\hbara}~d(p,q) v_+\\
\pp(v_-,t)& = & -\frac{p}{ m}\pp(v_-,q)-b_1\pp(v_-,p)+\frac{i}{\hbara}~d(p,q) v_-
\end{eqnarray}
with $$d(p,q):=  (b_0+\frac{1}{2}b_1 q).$$
These equations can be solved explicitly, e.g. with the method of characteristics. The solution corresponding to an initial condition
$$
v_{\pm}(p,q,t=0)=v^{(0)}_{\pm}(p,q)
$$
is given by
\begin{equation}
v_{\pm}(p,q,t)=e^{-i \delta_{\pm}(p,q,t)} v^{(0)}_{\pm}(p\pm b_1 t~,~ q-\frac{p}{ m}t\mp\frac{b_1}{ 2m} t^2)
\end{equation}
where $\delta_{\pm}(p,q,t)$ is a phase given by
$$
\hbar \delta_{\pm}=\pm\Bigl[qt-\frac{p}{2m}t^2\mp \frac{b_1}{6m}t^3  \Bigr]
\frac{b_1}{2}\pm b_0 t.
$$

\subsection{Physical interpretation -- classical-quantum entanglement}
The physical interpretation of $\psi=({v_+(p,q,t)},{v_-(p,q,t)})$ is that 
$|v_{\pm}(p,q,t)|^2$ gives the probability for the center of mass of the particle to have  a position $q$, a momentum $p$ and a spin component $\pm\hbar/2$ in the $q_3$-direction.
This dynamics can be interpreted as follows. 

We consider initial conditions that are a tensor product of the spin and the momentum-position, i.e. of the form
$$
\psi(t=0) =\xi^{(0)}(p,q)\otimes\dbinom{s_+}{s_-},
$$
with $|s_+|^2+|s_-|^2=1$.
Thus, in the initial state the spin and the center of mass are not entangled.
We consider an initial state that is well-localized both in momentum $p$ and in position $q$, described e.g. by narrow Gaussians
$$
\xi^{(0)}(p,q)=\frac{1}{\sqrt{\pi w_p w_q}} e^{-\frac{(p-p_0)^2}{2w_p^2}} e^{-\frac{(q-q_0)^2}{2w_q^2}}
$$
with $p_0=0$.\hfil The limit of complete localization corresponds to
$
\lim_{w_p,w_q\to 0}|\xi^{(0)}(p,q)|^2=  \delta(p-p_0)\delta(q-q_0).
$

We consider the following four examples of initial conditions: 
\begin{eqnarray*}
(i):\phantom{v}\qquad\psi(t=0) &=&\xi^{(0)}(p,q)\otimes\dbinom{1}{0}\\
(ii):\phantom{i} \qquad\psi(t=0) &=&\xi^{(0)}(p,q)\otimes\dbinom{0}{1}\\
(iii): \qquad\psi(t=0) &=&\xi^{(0)}(p,q)\otimes\dbinom{1}{1}\frac{1}{\sqrt{2}}\\
(iv): \qquad\psi(t=0) &=&\xi^{(0)}(p,q)\otimes\dbinom{s_+}{s_-}.
\end{eqnarray*}
In (i) and (ii) the wave packet of the center of mass  will follow a single trajectory without modification of its shape. It will move and accelerate in the direction $\pm q_3$, depending on the initial sign of the spin. \\
In cases (iii) and (iv) the state becomes a coherent superposition of two wave packets for the center of mass, that move in opposite directions. Both packets have the same shape, but the weight is given by a multiplicative factor $|s_{\pm}|^2$ determined by the initial state of the spin. We remark that in this case the spin and the center of mass are entangled, since  the state
 $\psi(t)$ cannot be written as a tensor product of a spin state $(s_1,s_2)$ and a center of mass state $\xi(p,q)$:
  $$\psi(t)=v_+(p,q,t)\otimes \binom{1}{0}+v_-(p,q,t)\otimes \binom{0}{1}\not=
  \xi(p,q,t)\otimes \dbinom{s_1}{s_2},$$
  since this would imply $v_+(p,q,t)=\beta~ v_-(p,q,t)$, with a constant $\beta$.\\
  This example shows that it is possible to entangle a classical and a quantum degree of freedom.


\section{\label{geometricQ-sect} Berezin-Toeplitz Quantization -- Geometric Quantization }

In this section we review, with a minimum of mathematical formalism, some of the main ideas of 
Berezin-Toeplitz quantization and its relation with geometric quantization.
We also summarize some elements of the theory of coherent states and their application to quantization.

In order to establish the relation between a quantum system and its classical counterpart we have to consider separately the relations between the Hilbert spaces, between the algebras of observables and between the derivations defining the dynamics. 

The Berezin-Toeplitz quantization consists of selecting a subspace
$\breve{\L}\subset \L_K$ of the Koopman Hilbert space and a map that assigns to each
multiplication operator $M_f$ on $\L_{K}$ an operator $T_{M_{f}}$ on $\breve{\L}$
defined as the projection of $M_f$ on  $\breve{\L}$.
In general two operators $T_{M_{f_{1}}},T_{M_{f_{2}}}$ corresponding to two different
multiplication operators $M_{f_{1}},M_{f_{2}}$ do not commute with
each other. \\
\\
Geometric quantization can be viewed as an extension of this procedure to the quantization of the generators of the dynamics, i.e. the Hamiltonians. The link between the quantization of the observables and of the  generators of the dynamics is given by the Tuynman relation \cite{Tuynman_JMP87}.

\subsection{Selection of a polarization subspace} 

As a first example we consider 
a system with one degree of freedom and phase space $\M=\RR^2$. The Koopman Hilbert space is
$\L_K=L_2(\RR^2,dp~dq)$, i.e. the square-integrable functions on the phase space. 

The first step in the procedure of quantization  is known in the literature on geometric quantization as the choice of a {\it ``polarization''}, which here we formulate as the choice of a {\it polarization subspace} $\breve\L\subset\L_K$. We present a simple description of  the construction in terms of action-angle variables $I,\theta$, defined by the canonical transformation
\begin{eqnarray} \label{beta-0}
 I&=\frac{1}{2}(p^2/{\beta_0}+ q^2\beta_0), \qquad &p=\sqrt{2I\beta_0}\sin\theta, \\
 \theta&=\arctan(p/\sqrt{\beta_0},q\sqrt{\beta_0}), \qquad &q=\sqrt{2I/\beta_0}\cos\theta, 
\end{eqnarray}
where ${\beta_0}$  is an arbitrary reference constant (with units of a mass times a frequency ${\beta_0}=m_0\omega_0$, or equivalently of an action times the square of a length).  
The function $\arctan(y,x)$ of two arguments is defined as the single-valued function that gives the unique angle $\theta\in [0,2\pi]$  such that 
$\cos\theta=y/\sqrt{y^2+x^2}$ and $\sin\theta=x/\sqrt{y^2+x^2}$. 

\subsubsection{Construction of a basis of $\L_K$} 

We will use the following basis of $L_2(\RR^2,dp~dq)$, expressed in action-angle coordinates:
\begin{equation}
\label{basis1}
\tilde\xi_{m',m}(I,\theta):= \nu_{m',m}~e^{- \frac{1}{2\alh}I} I^{(m'+m)/2} e^{i(m-m')\theta},\qquad m,m'\in \NN_0=\{0,1,2,\ldots\}
\end{equation}
where $ \nu_{m',m}$ is the normalization factor, and $\alh$ is an arbitrary fixed real constant. Since the argument of the exponential should be dimensionless, $\alh$ must have the dimension of an action. We remark that a constant that makes the variable dimensionless in the function $ I^{(m'+m)/2}$ is included in the normalization factor $\nu_{m',m}$, to simplify the notation.\\ \\
One can verify that (\ref{basis1})  is a basis of $L_2(\RR^2,dp~dq)=L_2(\RR,dp)\otimes L_2(\RR,dq)$ by first considering the known basis of $L_2(\RR,dp)$
$$
H_{m'}(p/{\sqrt{2\alh\beta_0}}) e^{- \frac{1}{2\alh\beta_0}p^2/2}, \qquad m'\in \NN_0
$$
where $H_{m'}$ are the Hermite polynomials, and the basis of $L_2(\RR,dq)$
$$
H_{m}(q\sqrt{{\beta_0}/({2\alh})}) e^{- \frac{\beta_0}{2\alh}q^2/2}, \qquad m\in \NN_0.
$$
 The set of functions
$$
p^{m'} q^m e^{- \frac{1}{2\alh}\bigl(p^2/\beta_0+ q^2\beta_0\big)/2}  \qquad m,m'\in \NN_0
$$
is therefore a basis $L_2(\RR^2,dp~dq)$, and defining the dimensionless  complex variable
\begin{equation}\label{zcoord}
z:= \frac{1}{\sqrt{2\alh}} (q\sqrt{\beta_0}+ip/\sqrt{\beta_0})=:\sqrt{\frac{I}{\alh}}e^{i\theta},
\end{equation}
another basis is given by the functions 
$$
\tilde\xi_{m',m}=\tilde\nu_{m',m}~e^{-\frac{1}{2} |z|^2}~z^{*m'} z^{m} 
\equiv \nu_{m',m}~ e^{- \frac{1}{2\alh}I} I^{\frac{m'+m}{2}} e^{i(m-m')\theta},
$$
which coincides with  (\ref{basis1}). 
We remark that the measure $d\mu$ expressed in the complex variables $z$ is
\begin{equation}
\label{ zmeasure}
d\mu(\uz):=dq~dp=2\lambda d^2z, \qquad{\rm with} \quad d^2z=dz_r~dz_i, \qquad z=z_r+iz_i.
\end{equation}

We can relabel the basis vectors through
\begin{eqnarray}
k & := & m-m'\qquad\qquad\qquad m=\frac{1}{2}(n+k) \\
n &: = & m+m'   \qquad\qquad\qquad m'=\frac{1}{2}(n-k)
\end{eqnarray}
and write the basis as
\begin{equation}
\label{basis2}
\xi_{n,k}(I,\theta):= \nu_{n,k} e^{-\frac{1}{2\alh}I} I^{n/2} e^{ik\theta},
\end{equation}
with $n\in\NN_0,\qquad
k\in \{-n,-n+2, -n+4,\ldots, n-2,n\} $, \qquad $\nu_n=(n!~2\pi~ \lambda^{n+1})^{-1/2}$. 

\subsubsection{Selection of a polarization subspace $\breve\L\subset\L_K$ by the choice of a subset of the basis} 

The selection of the polarization subspace can be performed by selecting a subset of this basis. In the standard Berezin quantization one selects the subspace $\breve{\L}\subset \L_K$ as follows:\\
First one chooses a particular value for the constant  $\alh=\hbarb$, where $\hbarb$ is equal to the constant that we had introduced in Eq. (\ref{koopman-phase}), when we introduced a phase in the unitary Koopman evolution. Then one selects the subspace
 generated by the subset $\{ \eta_n\}\subset\{\xi_{n,k}\}$  of the basis functions defined as
 \begin{eqnarray}\label{basis-polarization-subspace}
\eta_n(I,\theta):=\xi_{n,k=-n}(I,\theta)&\equiv& \nu_n~e^{-\frac{1}{2\hbarb}I}I^{n/2}e^{-in\theta} 
\qquad  {n\in\NN_0}\\
&\equiv&\tilde\nu_n~z^{*n}e^{-zz^*/2},
\end{eqnarray}
with $\nu_n=(n!~2\pi~ \hbarb^{n+1})^{-1/2}$ and $\tilde\nu_n=(n!~2\pi~ \hbarb)^{-1/2}$.\\
We remark that different choices of the constant $\alh$ lead to different subspaces, e.g. the function
$e^{-\frac{1}{2\alh'}I}$ is not contained in the subspace $\breve{\L}$ if $\alh'\neq \hbarb$. Thus the choice of $\alh$ equal to Planck's constant $\hbarb$ is non-trivial, in the sense that it is not just a conventional choice of units, but it is  an essential ingredient in the definition of the quantum model. Its numerical value in any given system of units must be determined by comparison with experiments.

\subsubsection{Definition of an isomorphism between $\breve\L\subset\L_K$ and Fock space}
 
The basis (\ref{basis-polarization-subspace}) of $\breve{\L}$ is labeled by a single index $n\in\NN_0$. One can define
an isomorphism $\Xi$ between the subspace $\breve{\L}$ (whose elements are functions of $p,q$) and a Hilbert space $\H$, that will be the Hilbert space of the quantum system, which can be defined formally as the space generated by a set of orthonormal states 
$\{|n\rangle\}_{n\in\NN_0}$.
The isomorphism is defined by
$$
\Xi : \eta_{n} \mapsto  |n\rangle
$$
which in the Dirac notation can be written as
$$
\Xi := \sum_n   |n\rangle \langle \eta_{n}|.
$$
As concrete examples for the quantum Hilbert space $\H$ we consider two examples: \\ \\
(i) We can take $\H$ as  the abstract Fock space $\F$ constructed from a ground state $|0\rangle$ and the creation operator $a^{\dag}$ : $|n\rangle:=  \nu_n{(a^{\dag})}^n |0\rangle$.\\ \\
(ii)
 One can take  $\H=L_2(\RR,dx)$
and for  $ |n\rangle$  the basis of eigenfunctions of the Hamiltonian of a harmonic oscillator
$$
H_{h.o.}:= -\frac{\hbar^2}{2m}\frac{d^2}{dx^2}+\frac{m\omega^2}{2}x^2,
$$
with $m$ and $\omega$ such that $m\omega=\beta_0$, where ${\beta_0}$ is the constant used in (\ref{beta-0}), i.e. 
\begin{equation}
\label{hosc-map}
\Xi : \eta_{n} \mapsto  |n\rangle=\varphi_n(x)= \nu_n H_n(x\sqrt{m\omega/\hbar})e^{-\frac{1}{2}x^2 m\omega/\hbar},
\end{equation}
where $H_n$ are the Hermite polynomials.

{\bf Remark: } Since $z^*:=\sqrt{\frac{I}{\hbarb}}e^{-i\theta}$, by Eq. (\ref{basis2}) the subspace $\breve{\L}$ can be also identified as being isomorphic to the Hilbert space of anti-holomorphic functions $g(z^*)$ \cite{Bargmann-1961,Bargmann-1962,Hall}, with scalar product $\langle g_1,g_2\rangle:=\int_{\M}dz~ e^{-|z|^2} g^*_1(z^*)~g_2(z^*)$, which is the usual formulation in the literature on geometric and Berezin quantization.
\\ \\

{\bf Summary:} The polarization subspace $\breve{\L}\subset \L_K$ is the subspace generated by the  orthonormal  set of functions
\begin{equation} \label{polarization-basis}
\eta_n:=\nu_n~I^{n/2}e^{-in\theta} e^{-\frac{1}{2\hbarb}I}\equiv \tilde\nu_n~z^{*n}e^{-zz^*/2}
\qquad  {n\in\NN_0},
\end{equation}
with $\nu_n=(n!~2\pi~ \hbarb^{n+1})^{-1/2}$ and $\tilde\nu_n=(n!~2\pi~ \hbarb)^{-1/2}$. The isomorphism $ \Xi : \eta_{n} \mapsto  |n\rangle$ gives the representation in the 
quantum Hilbert space $\H$.

\subsection{Toeplitz quantization of the observables}

The Toeplitz quantization of operators of the classical Hilbert space consists simply of 
taking the projection of the operator on the polarization subspace:
To a multiplication operator $M_f$ acting on $\L_K$ one associates an operator $T_{M_{f}}$
\begin{equation}
\label{toeplitz_operator}
M_f \mapsto T_{M_{f}}:= P_{\breve{\L}} M_f P_{\breve{\L}},
\end{equation}
where $P_{\breve{\L}} $ is the orthogonal projection from $\L_K$ to the polarization  subspace $\breve{\L}$. By composition with the isomorphism $\Xi$ one defines the associated operator on $\H$:
 $$
M_f \mapsto \widehat f\equiv\widehat T_{f}:= \Xi ~T_{f} ~\Xi^{-1}=
\Xi ~ P_{\breve{\L}} M_f  P_{\breve{\L}} ~\Xi^{-1},
$$
which can be expressed in terms of the bases $\xi_{n,k=-n}$ of $\breve{\L}$ and 
$|n\rangle$ of $\H$ as
\begin{equation}
\label{}
M_f \mapsto \widehat f\equiv \widehat T_{f}:=\sum_{n',n} |n'\rangle\langle\eta_{n'}|M_f |\eta_{n}\rangle\langle n|.
\end{equation}

One can calculate the matrix elements for some of the basic polynomial functions explicitly:
\begin{eqnarray}
\langle\eta_{n'}|z |\eta_{n}\rangle & = & \delta _{n',n-1}\sqrt{n}\\
\langle\eta_{n'}|z^* |\eta_{n}\rangle & = & \delta _{n',n+1}\sqrt{n+1}
\end{eqnarray}
and
\begin{eqnarray*}
&&\langle\eta_{n'}|z^m (z^*)^k |\eta_{n}\rangle 
 = \delta _{n',n-m+k} \times \\
&&\times
\sqrt{(n+k-m+1)\ldots (n+k-1)(n+k)~(n+k)(n+k-1)\ldots(n+1) },
\end{eqnarray*}
which leads to their identification in terms of creation-annihillation operators $a^{\dag}, a$ (either in the abstract Fock space $\F$ or in $L_2(\RR,dx)$:
\begin{eqnarray}
\widehat{z} & = &\hat a   \\
\widehat{z^*} & = &\hat a^{\dag}\\
\widehat{T}_{z^k z^{*m} }\equiv \widehat{z^k z^{*m} }& = &{\hat a}^k ~{\hat {a}}^{\dag m}. \label{antinormal}
\end{eqnarray}
We remark  that the Berezin-Toeplitz quantization with the chosen polarization subspace yields the operators  (\ref{antinormal}) in anti-normal ordering, i.e. with all the $\hat a^{\dag}$ on the right.

If we use the representation  $\H=L_2(\RR,dx)$ defined by the isomorphism (\ref{hosc-map}),
and 
\begin{eqnarray}
q & =  \sqrt{\frac{\hbar}{2\beta_0}} (z+z^*),  \qquad p  = & \sqrt{\frac{\hbar\beta_0}{2}} (z-z^*) \\
\hat x  & =  \sqrt{\frac{\hbar}{2\beta_0}} (\hat a+\hat a^{\dag}),  \qquad \hat p  = & \sqrt{\frac{\hbar\beta_0}{2}}  (\hat a+\hat a^{\dag})
\end{eqnarray}
 we obtain
\begin{eqnarray}
\widehat q & = & M_x=:\hat x\qquad ({\rm multiplication~by~} x )\\
\widehat p & = & -i\hbarb\pp( ,x) =:\hat p\\
\widehat {\phantom{i}p^2\phantom{i}} & = & (\widehat p)^2+\frac{\hbarb \beta_0}{2}\id= -\hbarb^2\frac{\partial^2}{\partial x^2} +\frac{\hbarb \beta_0}{2}\id\\
\widehat {\phantom{i}q^2\phantom{i}} & = &    (\widehat q)^2+\frac{\hbarb}{2 \beta_0}\id                            
= {\hat x^2}+\frac{\hbarb}{2 \beta_0}\id \\
\widehat{I} & = &    \frac{1}{2}\bigl( \frac{1}{\beta_0} {\hat p}^2    +  \beta_0 {\hat x}^2 \bigr) =-i\hbarb\pp( ,\theta)    \\              
\widehat{H_{h.o.}} =\omega \widehat{I} & = &  - \frac{\hbarb^2}{2m} \frac{\partial^2}{\partial x^2}   +   \frac{m\omega^2}{2}   {\hat x}^2   + \frac{\hbarb \omega}{2},       
\end{eqnarray}
where $H_{h.o.}  =  \omega I = \omega \hbar z z^* =\frac{1}{2m}  {p}^2    +   \frac{m\omega^2}{2}   {q}^2  $.
\subsection{\label{quantization-of-dynamics}Toeplitz quantization of the generators of the dynamics 
-- geometric quantization}

The Toeplitz quantization, that we first have defined for multiplication operators as the projection into the polarization subspace $\breve{\L}$,  can be extended to 
 the differential operators of the generators of  the dynamics:
 \begin{equation}
\label{toeplitz_operator_G}
G_f \mapsto T_{G_{f}}:= P_{\breve{\L}} G_f P_{\breve{\L}}.
\end{equation}
By composition with the isomorphism $\Xi$ one defines the associated operator on
the quantum Hilbert space $\H$:
 $$
G_g \mapsto\widehat{G_g} := \Xi ~P_{\breve{\L}} G_g P_{\breve{\L}} ~\Xi^{-1},
$$
which can be expressed in terms of the bases $\eta_{n}$ of $\breve{\L}$ and 
$|n\rangle$ of $\H$ as
\begin{equation}
\label{}
G_g \mapsto\widehat{G_g}=\sum_{n',n} |n'\rangle\langle\eta_{n'}|G_g |\eta_{n}\rangle\langle n|.
\end{equation}
The Schr\"odinger equation in the Hilbert space $\H$, corresponding to a classical Hamilton function $H_{cl}$ is thus given by
$$
i \pp(\psi,t) = \widehat{G}_{H_{cl}} \psi,\qquad
{\rm i.e.}\qquad i\hbar \pp(\psi,t) = \hat{H}_{H_{cl}}\psi, 
\qquad{\rm with}\qquad \hat{H}_{H_{cl}}:=\hbar \widehat{G}_{H_{cl}}.
$$

The Poisson brackets can be expressed in terms of the complex coordinates (\ref{zcoord}), choosing $\lambda=\hbarb$, as
$$
\bigl\{h,f\bigr\}:=\frac{i}{\hbarb}\sum_j \frac{\partial h}{\partial z_j}\frac{\partial f}{\partial z^*_j}-\frac{\partial h}{\partial z^*_j}\frac{\partial f}{\partial z_j},
$$
and $\AA_f$ as 
$$
\AA_f= -\frac{1}{2} \sum_j z\pp(f,z) + z^*\pp(f,z^*). 
$$
 For  a Hamilton function of the  form  $f=z^kz^{*m}$  the quantized generator of the dynamics is given by
 \begin{equation}
\label{Gquant}
\hbar\widehat{G_f}= \hat a^k \hat a^{\dag m} -km~\hat a^{(k-1)} \hat a^{\dag (m-1)}.
\end{equation}
This result can be obtained by the following steps: 
We first determine
\begin{eqnarray}
  \AA_f&=& -\frac{k+m}{2} z^k z^{* m} \\
\hbarb  \widetilde{X_f} &=& k z^{(k-1)} z^{* m} \pp( ,z^*) - m  z^{k} z^{* (m-1)} \pp( ,z) \label{Xlab}\\
\hbar G_f  &=& f +\AA_f+\hbar \tilde{X}_f \\
&=& \bigl( 1-\frac{k+m}{2}\bigr) z^k z^{* m}+k z^{(k-1)} z^{* m} \pp( ,z^*) - m  z^{k} z^{* (m-1)} \pp( ,z).\label{Glab}
\end{eqnarray}
Then we determine 
$
\widehat{X_f} := \Xi ~P_{\breve{\L}} \widetilde{X}_f P_{\breve{\L}} ~\Xi^{-1},
$ as
\begin{equation}
\label{}
\hbar\widehat{X_f} =\frac{k+m}{2} \hat a^k \hat a^{\dag m} -km~\hat a^{(k-1)} \hat a^{\dag (m-1)},
\end{equation}
which combined with 
 \begin{equation}
\widehat{T}_{z^kz^{*m}}= \hat a^k \hat a^{\dag m} 
\end{equation}
yields (\ref{Gquant}).
\\ \\
{\bf Remark:} Eq. (\ref{Gquant}) is a special case  of Tuynman's relation \cite{Tuynman_JMP87}, which allows to express the quantization of the generator of the dynamics $G_f$ in terms of the Toeplitz quantization of an associated observable $\tau(f)$:
\\ \\
{\bf Proposition:} (Tuynman's relation \cite{Tuynman_JMP87}) 
 \begin{equation} \label{Tuynman-relation}
\hbar\widehat{G}_{f}= \widehat{T}_{\tau(f)}, \quad {\rm with }\quad \tau(f):= f-\frac{\partial^2 f}{\partial z \partial z^*}.
\end{equation}
This relation can be written in a more general context as  $\tau(f):= f+\frac{\hbar}{4}\Delta_{dR}$, where 
$\Delta_{dR}$ is the de Rham Laplacian, which in our case is given by $\Delta_{dR}= -\frac{4}{\hbar}\frac{\partial^2}{\partial z \partial z^*}$. We notice that in  \cite{Tuynman_JMP87}  the complex variables $z$ are defined with a convention that differs from ours by a factor $\sqrt{2}$.
 We remark that 
\begin{equation}
\label{Laplacian}
\frac{\partial^2 f}{\partial z \partial z^*} =\frac{\hbar}{2}\biggl( \frac{1}{\beta_0}\frac{\partial^2 f}{\partial q^2 } 
+{\beta_0}\frac{\partial^2 f}{\partial p^2 }   \biggr).
\end{equation} 
In Table   \ref{table-G} we give  for some examples of the expressions  of  the operators $\AA_f, \widetilde{X}_f, \widehat{T}_f,
\widehat{G}_f$. We use the notation $\hat x =M_x; \quad\hat p=-i\hbarb\pp( ,x)$  . 
In Table \ref{table-G} we have expressed the quantized operator corresponding to the action $I$, in the phase representation \cite{Bialynicki-Birula-phase-PRA-1976,Bialynicki-Birula-phase-APP-1980,Guerin-Monti-JPA1997} where the Hilbert space $\H$ is generated by the functions $\{e^{in\theta}, \quad n\in \NN_0\}$. We remark that there is no simple explicit expression for the 
$\widehat{T}_V,\widehat{G}_V$ corresponding  to a general potential $V(q)$.
\\  
For the harmonic oscillator we have:
\begin{eqnarray}
H_{h.o.} &=&\omega \biggl(\frac{1}{2m\omega}p^2+\frac{1}{2}m\omega q^2\biggr) = \omega I=\hbar\omega zz^*\\
\AA_{H_{h.o.}}&=& - H_{h.o.}\\
\tilde X _{H_{h.o.}}&=& -i\omega \biggl(\frac{1}{2m\omega}p\pp( ,q) 
+\frac{1}{2}m\omega q\pp( ,p) \biggr)=-i\omega \pp( ,\theta)\\
G_{H_{h.o.}}&=&\tilde X _{H_{h.o.}}\\
\widehat{T}_{H_{h.o.}}&=&\omega \biggl(\frac{1}{2m\omega}\hat p^2 +\frac{1}{2}m\omega \hat x \biggr) + \frac{\hbar \omega}{2} \\
&=& \hbar\omega \hat a \hat a^{\dag} = \hbar\omega\bigl( \hat a^{\dag} \hat a +\frac{1}{2}\bigr) +\frac{\hbar\omega}{2}\\
\hat H_{h.o.}&\equiv&\hbar\widehat{G}_{H_{h.o.}}=\hbar\widehat{X}_{H_{h.o.}}\\&=& \hbar\omega (\hat a\hat a^{\dag} -1) = \hbar\omega \hat a^{\dag} \hat a \\
&=&  \biggl(\frac{1}{2m}\hat p^2 +\frac{1}{2}m\omega^2 \hat x \biggr) - \frac{\hbar \omega}{2}.
\end{eqnarray}
\\ \\
\centerline{
\begin{tabular}{|*{1}{c|}|*{3}{c|}|*{1}{c|}|*{1}{c|}}
\hline
$ $f & $\AA_f$ & $\tilde X_f$  & $\Gen_f$& $\widehat{T_f}$  &$\hbarb \widehat{G_f}=\hat H_f$\\ 
\hline
 $ 1$   &0  &0  &$\frac{1}{\hbara}\id$ &$\id$  &$\id$\\   
\hline
$ z^k z^{*m} $ &$-\frac{k+m}{2} z^k z^{* m} $  &(\ref{Xlab}) &(\ref{Glab}) &
$ {\hat a}^k {\hat a}^{\dag m}$&
$  {\hat a}^k {\hat a}^{\dag m}-km~{\hat a}^{(k-1)} {\hat a}^{\dag (m-1)} $\\
\hline
  $q$ &$-\frac{1}{2} q$  & $i\pp( ,p)$  &$\frac{1}{2\hbara} q +i\pp( ,p)$&$\hat x$ &$\hat x$ \\ 
   \hline
     $p$ &$-\frac{1}{2} p$  & $-i\pp( ,q)$  &$\frac{1}{2\hbara} p -i\pp( ,q)$&$\hat p$ &$\hat p$\\ 
   \hline
    $q^2$ &$-q^2$  &$i2q\pp( ,p)$  &$i2q\pp( ,p)$  &${\hat x}^2+\frac{\hbarb}{2\beta_0}$ &${\hat x}^2-\frac{\hbarb}{ 2\beta_0}$\\
\hline
  $p^2$ &$-p^2$  & $-i2p\pp( ,q)$  &$-i2p\pp( ,q)$   &${\hat p}^2+\frac{\hbarb \beta_0}{ 2}$ &${\hat p}^2-\frac{\hbarb \beta_0}{ 2}$\\
\hline
 $ I$& $ -I$ & $-i\{I,\cdot\}$ &$-i\{I,\cdot\}  \equiv -i\pp( ,\theta)$  &$-i \hbarb\pp( ,\theta)+\hbarb\id $&$-i \hbar\pp( ,\theta)$\\
\hline
$V(q)$& $-\frac{1}{2}q\pp(V,q)$ &$i\pp(V,q)\pp( ,p)$  &$\frac{1}{\hbara}V-\pp(V,q)\bigl(\frac{q}{2\hbara}  -i\pp( ,p)\bigr)$   & &\\
\hline
\end{tabular}
\label{table-G}
}
\\ \\
Table  \ref{table-G}: Some examples of the expressions  of  the operators $\AA_f, \widetilde{X}_f, \widehat{T}_f,
\widehat{G}_f$. 

\newpage

\subsection{Quantization by coherent states}

An alternative formulation of  the quantization of multiplication operators on $\L_K$  using coherent states was proposed in Refs. \cite{Ali,Gazeau-quantization-CS,Gazeau-quantization-CS-exemples,Gazeau-fuzzy-sphere-JPA-2007}. As we will  discuss below, this formulation yields the same quantized operators as the Toeplitz quantization. It can also be extended to yield the same quantization of the generators of the dynamics.

\subsubsection{Definition of coherent states}

There are several definitions of coherent states that emphasize different types of properties 
\cite{Klauder-CS-2001,Klauder-Skagerstam-book,Ali-Antoine-Gazeau-book-2000}: Minimization of the Heisenberg uncertainty relations \cite{Barut-Girardello-CMP-1972}, group theoretical properties 
\cite{Zhang-Feng-Gilmore-RMP,Perelomov-book}, annihilation operator eigenfunctions 
\cite{Nieto-Simmons-PRL-78}.

For the purpose of establishing relations between classical and quantum systems one can use a definition that addresses only one property that is shared by all the other definitions: We consider a Hilbert space $\H$ and a phase space $\M$ with a volume measure $d\mu$.

{\bf Definition:} A {\it complete set of vectors indexed by the points of $\M$},  $\{|\zeta^{\H}_{\uzo}\rangle\in\H \}_{\uzo\in\M}$, is defined by a continuous map $\M\to \H$, 
$\uzo \mapsto |\zeta_{\uzo}\rangle$, such that
\begin{equation}
\label{completenessa}
\int_{\M}d\mu(\uzo)~  | \zeta_{\uzo}^{\H}\rangle  \langle \zeta_{\uzo}^{\H} | =\id_{\H}.
\end{equation}

{\bf Definition:} The corresponding {\it  coherent states} are defined as the normalized vectors
\begin{equation}
|C_{\uzo}^{\H}\rangle := \frac{1}{\nu(\uzo)}  | \zeta_{\uzo}^{\H}\rangle,
\end{equation}
with $\nu^2(\uzo):=  \langle \zeta_{\uzo}^{\H}  | \zeta_{\uzo}^{\H}\rangle$. The non-normalized vectors $|\zeta^{\H}_{\uzo}\rangle$,  will also be  called {\it unnormalized coherent states}. 

\subsubsection{Construction of the coherent states determined by the selection of a polarization subspace}

In Refs. \cite{Ali,Gazeau-quantization-CS,Ali-Antoine-Gazeau-book-2000} a general construction of coherent states was proposed which is based on the selection of a  polarization subspace $\breve{\L}$. 
We assume that the elements of $\breve{\L}$ are continuous functions. 
One defines for each point $\uzo\in\M$ of the phase space  an {\it evaluation functional}, that assigns to each function its value at the point $\uzo$:
\begin{eqnarray}
\delta_{\uzo} : \breve{\L} & \to & \CC \\
 |\xi\rangle& \mapsto& \xi(\uzo).
\end{eqnarray}
Under the assumption that this linear map is continuous, by Riesz's theorem \cite{Reed-Simon-1} there is a unique vector
$|\zeta_{\uzo}\rangle\in \H$ such that $\forall |\xi\rangle\in\breve{\L}$
$$
\delta_{\uzo} |\xi\rangle = \langle \zeta_{\uzo}~|~ \xi\rangle.
$$
One can give an explicit expression of the functional $ \langle\zeta_{\uzo}|$ in terms of the arbitrary orthonormal basis of continuous functions$\{ |\eta_n\rangle \}\in \breve{\L}$:
\begin{equation}
\label{evaluationfunctional}
\delta_{\uzo}\equiv \langle \zeta_{\uzo}| = \sum_n \eta_n(\uzo) \langle \eta_n |,
\end{equation}
since, $\forall |\xi\rangle\in\breve{\L}$,
$
|\xi\rangle= \sum_n | \eta_n\rangle \langle \eta_n |  \xi\rangle   
$
and thus,
$$
\delta_{\uzo} |\xi\rangle =\sum_n  \delta_{\uzo}(| \eta_n\rangle) 
\langle \eta_n |  \xi\rangle\\
=\sum_n  \eta_n(\uzo) \langle \eta_n |  \xi\rangle.
$$
The corresponding vector can thus be written as
\begin{equation}
\label{evaluationvector}
| \zeta_{\uzo}\rangle = \sum_n \eta^*_n(\uzo) | \eta_n \rangle.
\end{equation}
The set of  vectors $| \zeta_{\uzo}\rangle$ satisfies the following completeness relation:
\begin{equation}
\label{completeness}
\int_{\M}d\mu(\uzo)~ | \zeta_{\uzo}\rangle\langle \zeta_{\uzo}| = \id_{\breve{\L}},
\end{equation}
since 
\begin{eqnarray*}
\int_{\M}d\mu(\uzo) | \zeta_{\uzo}\rangle\langle \zeta_{\uzo}| &=&
 \sum_{n',n}\Biggl( \int_{\M}d\mu(\uzo)~  \eta^*_{n'}(\uzo) \eta_{n}(\uzo\Biggr) | \eta_{n'} \rangle  \langle \eta_n | \\
  &=&
 \sum_{n',n} \delta_{n',n} | \eta_{n'} \rangle  \langle \eta_n | =\id_{\breve{\L}},
\end{eqnarray*}
where we have used $ \int_{\M}d\mu(\uzo)~  \eta^*_{n'}(\uzo) \eta_n(\uzo)=\delta_{n',n}$.\\
We remark that, since the evaluation vector $| \zeta_{\uzo}\rangle$ is a function  of $\uz$ which belongs to  the subspace $\breve{\L}$,
we can write it as
$$
\zeta_{\uzo}(\uz)= \sum_n \eta^*_n(\uz_0) ~ \eta_n (\uz).
$$
In the limit case when the polarization subspace coincides with the total Hilbert space $\breve{\L}=\L_K$,
the evaluation vector tends formally to a Dirac delta function:
$$
\zeta_{\uzo}(\uz) \xrightarrow[{\H\to L_K}]{} \delta(\uz-\uzo)
$$

The {\it  coherent states} $|C_{\uzo}\rangle$ determined by the choice of the subspace $\breve{\L}$ are defined by normalizing the vectors $| \zeta_{\uzo} \rangle$:
$$
|C_{\uzo}\rangle := \frac{1}{\nu(\uzo)}| \zeta_{\uzo} \rangle, $$
with
$$
 \nu^2(\uzo):=\langle \zeta_{\uzo} | \zeta_{\uzo} \rangle =
\sum_{n',n}      \eta_{n'}(\uzo) \eta^*_n(\uzo) \langle \eta_{n'}| \eta_n \rangle
=\sum_n|\eta_n(\uzo)|^2.
$$

One can define the analogue of the the evaluation vectors and of the
evaluation functionals  in  the quantum Hilbert space $\H$ (e.g. in Fock space $\F$ or in $L_2(\RR,~dx$)) by
\begin{eqnarray}
\label{evaluation-Fock}
| \zeta_{\uzo}^{\H}\rangle  & := & \Xi\biggl(| \zeta_{\uzo}\rangle  \biggr) = \sum_n \eta^*_n(\uzo) | n \rangle \\
\label{CS-Fock}
\delta_{\uzo}^{\H}\equiv \langle \zeta_{\uzo}^{\H}|  
& := & \sum_n \eta_n(\uzo) \langle n |,
\end{eqnarray}
where $\{| n\rangle\}$ is the orthogonal basis of the space $\H$.
\\ \\
{\bf Remark :} As we will see in Section (\ref{quantization-observables-sect}) in the quantization of observables and of the dynamics we don't actually use the normalized coherent states $|C_{\uzo}\rangle$, but directly the {\it evaluation functionals } $\langle \zeta_{\uzo} |$, $\langle \zeta_{\uzo}^{\H} |$ and their duals, the {\it evaluation vectors}  $| \zeta_{\uzo}\rangle$, $| \zeta_{\uzo}^{\H}\rangle$ (that we will also call
{\it unnormalized coherent states}). The essential property are the completeness relations (\ref{completenessa}),(\ref{completeness}). We remark, however, that for the standard Glauber coherent states \cite{Schroedinger-CS-1926,Glauber-CS-1963}, as well as for the spin or atomic coherent states of Gilmore \cite{Gilmore-Mex-1974} and of Perelomov \cite{Perelomov-CMP-1972}, the normalization factor $\nu(\uzo)$ is independent of $\uzo$, and thus the coherent states and the unnormalized coherent states are related by just a multiplicative constant.
\\  \\
In summary, the choice of a polarization subspace $\breve{\L}$ and the isomorphism $\Xi$ define the coherent states  (\ref{CS-Fock}) uniquely. We will see in Section 
(\ref{section-reconstruction})
an inverse property: A given set of coherent states in $\H$ determines uniquely   a polarization subspace $\breve{\L}\subset\L_K$ and the isomorphism  $\Xi$.

\subsubsection{\label{quantization-observables-sect} Coherent state  quantization of observables}

Using these unnormalized coherent states one can associate to each multiplication operator $M_f$ on $\L_K$ an operator $C_{M_{f}}$ on the subspace  $\breve{\L}$:
\begin{equation}
\label{cs_quantization}
C_{M_{f}}:= \int_{\M}d\mu(\uzo)  f(\uzo) | \zeta_{\uzo}\rangle  \langle \zeta_{\uzo} |.
\end{equation}

The following argument shows that the quantized operator $C_{M_{f}}$ defined by Eq. (\ref{cs_quantization}) with the coherent states, is identical to the Toeplitz operator (\ref{toeplitz_operator}) defined by projection on the polarization subspace $\breve{\L}$: $C_{M_{f}}\equiv T_{M_{f}}$.\\
By inserting the representations (\ref{evaluationfunctional})(\ref{evaluationvector})
into (\ref{cs_quantization}), we obtain
\begin{eqnarray*}
C_{M_{f}} &=& \sum_{n',n} \int_{\M}d\mu(\uzo)~ f(\uzo)~ \eta^*_{n'}(\uzo)~  \eta_{n}(\uzo)~ |\eta_{n'}\rangle\langle \eta_{n}| \\
 &=& \sum_{n',n} |\eta_{n'}\rangle\Bigl( \int_{\M}d\mu(\uzo)~ f(\uzo)~ \eta^*_{n'}(\uzo) ~\eta_{n}(\uzo)\Bigr)~ \langle \eta_{n}|,
\end{eqnarray*}
and using the fact that
$$
\int_{\M}d\mu(\uz)~ f(\uz) ~\eta^*_{n'}(\uz) ~ \eta_{n}(\uz)
\equiv \langle  \eta_{n'} |M_f |\eta_{n}\rangle
$$
and defining the projector $P_{\breve{\L}}:=\sum_n |\eta_{n}\rangle\langle \eta_{n} |$ into the polarization subspace $\breve{\L}$, we can write
\begin{eqnarray*}
C_{M_{f}}& = & \sum_{n',n}  |\eta_{n'}\rangle\langle  \eta_{n'} |M_f |\eta_{n}\rangle\langle \eta_{n}| \\
& = & P_{\breve{\L}}M_fP_{\breve{\L}}\equiv T_{M_{f}},
\end{eqnarray*}
which is the Toeplitz operator (\ref{toeplitz_operator}).

Using the analogues of the evaluation vectors and functionals 
defined by Eqs (\ref{evaluation-Fock})
 in the quantum Hilbert space $\H$ (i.e. on Fock space $\F$ or in $L_2(\RR,dx)$),  the quantized operator $\hat f$ corresponding to the observable  $f$ is given by
\begin{equation}
\label{cs_quantization_fock}
\hat f= \int_{\M}d\mu(\uzo)~  f(\uzo) ~| \zeta_{\uzo}^{\H}\rangle 
 \langle \zeta_{\uzo}^{\H} |.
\end{equation}

\subsubsection{Coherent state  quantization of the generators of the dynamics}

The  Toeplitz quantization of the generator of the dynamics can also be expressed in terms of coherent states. The formula (\ref{cs_quantization}), originally defined for multiplication operators, can be extended to the differential operators $X_g$ appearing in the generators $G_g$

\begin{equation}
\label{cs_quantization_vf}
C_{X_g}:= \int_{\M}d\mu(\uzo)~  | \zeta_{\uzo}\rangle~ X_g^{\uzo} (\langle \zeta_{\uzo} |)
\end{equation}
where  the notation $X_g^{\uzo} $ indicates that the  differential operator acts on the variables $\uzo$ and not on $\uz$ :
$$
X_g^{\uzo} (\langle \zeta_{\uzo} |)=X_g^{\uzo} \sum_n\eta_n(\uzo) \langle\eta_n|
= \sum_n X_g(\eta_n(\uzo)) \langle\eta_n|=
\sum_n \{g(\uzo),\eta_n(\uzo)\} \langle\eta_n|.
$$
The coherent state quantization of the generators of the dynamics can thus be written as
\begin{equation}
\label{cs_quantization_G}
C_{G_g}:= \int_{\M}d\mu(\uzo)~  | \zeta_{\uzo}\rangle~ G_g^{\uzo} (\langle \zeta_{\uzo} |).
\end{equation}
The result is again the same one as the one obtained by Toeplitz quantization:
$$
C_{G_g}=T_{G_g}.
$$
This can ve verified by an argument along the same line as the one for the multiplication operators:
\begin{eqnarray}
C_{G_{g}} &=& \int_{\M}d\mu(\uzo)~  | \zeta_{\uzo}\rangle~ G_g^{\uzo} (\langle \zeta_{\uzo} |)\\
 &=& \sum_{n',n} \int_{\M}d\mu(\uzo)~  \eta^*_{n'}(\uzo)~G_g^{\uzo} ( \eta_{n}(\uzo))~ |\eta_{n'}\rangle\langle \eta_{n}| \\
 &=& \sum_{n',n} |\eta_{n'}\rangle\Bigl( \int_{\M}d\mu(\uzo)~  \eta^*_{n'}(\uzo) ~G_g^{\uzo} (\eta_{n}(\uzo))\Bigr)~ \langle \eta_{n}|\\
 &=& \sum_{n',n} |\eta_{n'}\rangle\langle \eta_{n'}~| ~G_g ~|\eta_{n}\rangle \langle \eta_{n}|\\
 & = & P_{\breve{\L}}G_g P_{\breve{\L}}\equiv T_{G_{g}}.
\end{eqnarray}

\section{\label{dequantization-section} Dequantization by coherent states }

Coherent states can be used for the opposite process,  
 called {\it dequantization}, which is the construction of a classical system for a given quantum  model 
 \cite{Gracia-Bondia-dequantization-1992,Ali-Antoine-dequantization-1994}. 
 
The general problem of dequantization can be formulated as follows: Given a quantum system defined on a Hilbert space $\H$, with observables $\hat A$ and  a dynamics generated by a Hamiltonian $\hat H$, the goal is to find
\\ \\
(a) a phase space manifold $\M$ and a measure $d\mu$,\\
(b) a subspace $\breve{\L}\subset L_2(\M, d\mu)$, and an isomorphism $\Xi$ between $\breve{\L}$ and $\H$,\\
(c) for each relevant observable $\hat A$ a  function $f_A: \M\to \C$ such that the Toeplitz quantization of $f_A$ yields the operator $A$: $\Xi~ T_{f_{A}} \Xi^{-1}= \hat A$,\\
(d) a function $H_{cl}$ such that the Toeplitz quantization of $G_{H_{cl}}$ yields the operator $\hat H$.
\\ \\
We remark that dequantization is not a classical limit procedure involving $\hbar\to\infty$ but a correpondence, i.e. a map that assigns a classical system to a given quantum system.
\\ \\
We remark that the dequantization of the Hamiltonian, i.e. of the generator of the quantum dynamics, is different than the dequantization of the observables.
A procedure of dequantization along the above requirements can be formulated using coherent states as follows.

\subsubsection{(a) Construction of a phase space manifold from a set of coherent states}

The first step is the construction of a set of coherent states. For any given quantum system, the choice of coherent states is not unique. The approach of Gilmore  
and of Perelomov,  
based on a group theoretical construction, yields  a phase space manifold $\M$, a measure $d\mu$ and a set of unnormalized coherent states
satisfying  the completeness relation $\int d\mu(\uzo)~|\zeta_{\uzo}\rangle\langle \zeta_{\uzo} | =\id $. This step is described in detail in Refs. \cite{Zhang-Feng-Gilmore-RMP,Perelomov-book,Zhang-Feng-PhysRep-95}. \\

\subsubsection{\label{section-reconstruction} (b) Construction of the  polarization subspace and of the isomorphism  $\Xi$  from a given set of unnormalized coherent states}

If a complete set of unnormalized coherent states $\{|\zeta^{\H}_{\uzo}\rangle\in\H \}_{\uzo\in\M}$ is given, one can construct a polarization subspace $\breve{\L}\subset\L_K$ and an isomorphism $\Xi:\breve{\L}\to\H$   such that the states $| \zeta_{\uzo}\rangle$ defined by Eq.
(\ref{evaluationvector}) coincide with the states $|\zeta^{\H}_{\uzo}\rangle$ :
$$
\Xi | \zeta_{\uzo}\rangle = |\zeta^{\H}_{\uzo}\rangle.
$$
This can be shown as follows. We introduce a map $\Xi_{Hus}: \H \to \L_K$ by
$$
\Xi_{Hus}: \psi \mapsto \xi\qquad{\rm defined~by}\quad \xi(\uzo):= \langle \zeta^{\H}_{\uzo}  |\psi\rangle.
$$
We will call $\Xi_{Hus}$ the {\it Husimi map} since $|\xi(\uzo)|^2$ is the Husimi function corresponding to the state $\psi$.
The Husimi map defines an isomorphism between $\H$ and a subspace $\breve{\L}$ of $\L_K$, that satisfies the following properties:\\
(i) $\Xi_{Hus}$ is a continuous linear map that preserves the scalar products, i.e.
$$
\langle~ \Xi_{Hus}\bigl(\psi_1\bigr)~|~\Xi_{Hus}\bigl(\psi_2\bigr)~\rangle_{\L_{K}} =  \langle\psi_1 | \psi_2 \rangle_{\H},
$$
since
\begin{eqnarray*}
\langle~ \Xi_{Hus}\bigl(\psi_1\bigr)~|~\Xi_{Hus}\bigl(\psi_2\bigr)~\rangle_{\L_{K}} & = & 
\int_{\M}d\mu(\uzo) ~\biggl( \Xi_{Hus}\bigl(\psi_1\bigr)\biggr)^* \Xi_{Hus}\bigl(\psi_2\bigr) \\
 & = &\int_{\M}d\mu(\uzo) ~\langle\psi_1| \zeta^{\H}_{\uzo} \rangle_{\H}\langle \zeta^{\H}_{\uzo} | \psi_2\rangle_{\H}\\
 & = &\langle\psi_1| \Bigl(\int_{\M}d\mu(\uzo) ~|\zeta^{\H}_{\uzo} 
  \rangle\langle \zeta^{\H}_{\uzo}|\Bigr) | \psi_2\rangle\\
  & = &\langle\psi_1| \psi_2\rangle_{\H}.
 \end{eqnarray*}
This implies that the image of $\psi\in\H$ is indeed in the space $\L_{K}$ of square-integrable functions.\\
(ii) The image $\Xi_{Hus}(\H) =:\breve{\L}$ is a subspace of $\L_{K}$.\\ \\
(iii) We chose an arbitrary orthonormal basis $\{|n\rangle\}_{n\in I\subset \ZZ}$ of $\H$. Its image by the Husimi map defines
$$
\eta_n:=\Xi_{Hus} |n\rangle.
$$
The set of functions $\{\eta_n\}_{n\in I\subset \ZZ}$ is an orthonormal set that spans the subspace 
$\breve{\L}\subset \L_{K}$. If we define the isomorphism
$\Xi:\breve{\L} \to \H$ by  $\eta_n \mapsto |n\rangle$, we can identify it as the inverse of the Husimi map : $\Xi=\Xi^{-1}_{Hus}$, and 
\begin{eqnarray*}
\Xi|\zeta_{\uzo}\rangle &=& \Xi \sum_{n}\eta^*_n(\uzo)~|\eta_n\rangle \\
&=& \sum_{n}\eta^*_n(\uzo)~|n\rangle =|\zeta^{\H}_{\uzo}\rangle.
 \end{eqnarray*}

  We consider as an example the case of the standard Glauber coherent states defined on the Fock space $\H$ by
 $$
|C_{\uz}^{\H}\rangle :=        e^{z\hat a^{\dag}-z^*\hat a} | n=0\rangle.
 $$
 With respect to the basis $\{ |n\rangle\}$ they are expressed as
 \begin{equation}\label{ }
|C_{\uz}^{\H}\rangle = e^{-zz^*/2}\sum_{n=0}^\infty
\frac{z^n}{\sqrt{n!}}| n\rangle.
\end{equation}
Since they satisfy the completeness relation
$$
\id=\frac{1}{\pi} \int d^2z |C_{\uz}^{\H}\rangle \langle C_{\uz}^{\H}|
=\frac{1}{2\pi\hbar} \int d\mu(\uz) |C_{\uz}^{\H}\rangle \langle C_{\uz}^{\H}|,
$$
the unnormalized coherent states are
$$
 |\zeta_{\uz}^{\H}\rangle=\frac{1}{\sqrt{2\pi\hbar}}  |C_{\uz}^{\H}\rangle.
$$
The image of the corresponding  Husimi map $\Xi_{Hus}: \H \to \L_K$ is spanned by the following vectors:
\begin{equation}
\label{ }
\Xi_{Hus} | n\rangle \equiv    \langle \zeta_{\uz}^{\H} | n\rangle=   \frac{1}{\sqrt{2\pi\hbar n!}}  z^n e^{-zz^*/2}.
\end{equation}
This subspace is different from the polarization subspace $\breve \L$ we chose in (\ref{polarization-basis}), since $z^n$ generates holomorphic instead of anti-holomorphic functions. In order to obtain the subspace $\breve \L$ we have to choose a slightly different set of coherent states, exchanging $z$ and $z^*$ :
\begin{equation}\label{conjugate-Glauber-CS}
|{C'}^{\H}_{\uz}\rangle :=        e^{z^*\hat a^{\dag}-z\hat a} | n=0\rangle,
\qquad  |{\zeta'}_{\uz}^{\H}\rangle=\frac{1}{\sqrt{2\pi\hbar}}  |{C'}_{\uz}^{\H}\rangle.
\end{equation}
 which leads to the subspace generated by
\begin{equation}
\label{ }
\Xi'_{Hus} | n\rangle \equiv    \langle{ \zeta'}^{\H}_{\uz} | n\rangle=   \frac{1}{\sqrt{2\pi\hbar n!}}  z^{*n} e^{-zz^*/2},
\end{equation}
which is equal to the polarization subspace $\breve\L\in L_K$ of the Berezin-Toeplitz quantization that we defined in (\ref{polarization-basis}).

\subsubsection{(c) Dequantization of the observables --  covariant and contravariant symbols}
 
 For a given operator $\hat A$ one can define two types of {\it symbols}, which are functions or more generally distributions on the phase space $\M$:\\
\\
(i)  The  {\it contravariant symbol} $f_{\hat A}$ -- also called {\it upper bound symbol} 
or {\it P-symbol} -- is defined as a 
function (or more generally distribution) such that 
$
\hat A = \widehat{T}_{f_{\hat A}} , 
$
i.e.
\begin{equation}\label{contravariant-symbol}
\hat A = \widehat{T}_{f_{\hat A}} \equiv \Xi ~ P_{\breve{\L}} f_{\hat A}  P_{\breve{\L}} ~\Xi^{-1}
\equiv  \int_{\M}d\mu(\uzo)  f_{\hat A}(\uzo)~ | \zeta^{\H}_{\uzo}\rangle  \langle \zeta^{\H}_{\uzo} |. 
\end{equation}
(ii) The  {\it covariant symbol} $S_{\hat A}$ -- also called {\it lower bound symbol} or {\it Q-symbol} -- is defined as
\begin{equation}\label{covariant-symbol}
 S_{\hat A}(\uzo):=   \langle \zeta^{\H}_{\uzo} |\hat A|\zeta^{\H}_{\uzo}\rangle.
\end{equation} 
 
 We remark that  while for some operators $\hat A$ the symbols can be expected to be smooth  functions on $\M$ for other operators the symbols may not be well-defined or they may be a more singular object like e.g. a distribution. The validity of the following formal relations between an operator $\hat A$ and its covariant and contravariant symbols must be analyzed for each particular type of operator.

 1) The covariant Q-symbol $S_{\hat A}(\uzo)$ can be calculated directly, provided that the coherent states are in the domain of definition of the operator $\hat A$, i.e. provided that the scalar product is well defined. For the standard Glauber coherent states, it can be expressed also as \cite{Zhang-Feng-Gilmore-RMP,Klauder-Skagerstam-book}
 \begin{equation}
\label{ }
 S_{\hat A}(z_0) = \frac{1}{\pi}\int_{\CC}d^2z~ e^{z_0z^*-z_0^*z}~ Tr\bigl(\hat A   e^{-z^* \hat a}  e^{z \hat a^{\dag}} \bigr).
\end{equation}

 2) The contravariant  P-symbol $ f_{\hat A}(z_0)$ can be written as the following formal expression: 
 \begin{equation}
\label{ }
 f_{\hat A}(z_0) =  \frac{1}{\pi}\int_{\CC}d^2z~ e^{z_0z^*-z_0^*z}~ Tr\bigl(\hat A e^{z \hat a^{\dag}}  e^{-z^* \hat a}  \bigr).
\end{equation}

The covariant symbol $S_{\hat A}$ is generally easier to calculate and more regular that the covariant one $ f_{\hat A}$. $ f_{\hat A}$  can be expressed in terms of $S_{\hat A}$  through their Fourier transforms : defining the Fourier transforms
 \begin{eqnarray}
\tilde f_{\hat A}(w)  & := & \frac{1}{ \pi} \int_{\CC} d^2z~ f_{\hat A}(z) e^{w z^*-w^* z}\\
\tilde S_{\hat A}(w) & := &  \frac{1}{\pi } \int_{\CC} d^2z~ S_{\hat A}(z) e^{w z^*-w^* z}
\end{eqnarray}
and their inverses
 \begin{eqnarray}
 f_{\hat A}(z) & := & \frac{1}{ \pi} \int_{\CC} d^2z~\tilde f_{\hat A}(w)  e^{-w z^*+w^* z}\\
 S_{\hat A}(z) & := & \frac{1}{ \pi} \int_{\CC} d^2z~\tilde S_{\hat A}(w)  e^{-w z^*+w^* z}\end{eqnarray}
one can establish \cite{Klauder-Skagerstam-book,Zhang-Feng-Gilmore-RMP} the relation
 $$
\tilde f_{\hat A}(w) =  e^{w^* w}  \tilde S_{\hat A}(w).
 $$
which, applying the inverse Fourier transform,  can be written as
  \begin{equation}\label{fAexp} 
 f_{\hat A}(z) =  e^{-\frac{\partial^2}{\partial z\partial z^*}}  S_{\hat A}(z) 
 =e^{-\frac{\partial^2}{\partial z\partial z^*}} \langle \zeta^{\H}_{z} |\hat A| \zeta^{\H}_{z} \rangle.
\end{equation}
 {\bf Remark:} This relation is also true for the conjugate Glauber coherent states (\ref{conjugate-Glauber-CS}), since it is invariant upon the exchange of $z$ an $z^*$.
 
 \subsubsection{(d) Dequantization of the Hamiltonian generator of the dynamics}
 
 Given an operator $\hat H$ in $\H$, we want to determine a function $H(p,q)$, such that 
 $$
\hbar \widehat{G}_H = \hat H.
 $$
 Using Tuynman's relation, this is equivalent to 
 $$
 \widehat{T}_{\tau(H)} = \hat H
 $$
The function $ h:=\tau(H)$ is, by definition, the contravariant symbol of $\hat H$, which according to Eqs. (\ref{covariant-symbol}) (\ref{fAexp}) can be expressed as
  \begin{equation} \label{hexp}
 h=  e^{-\frac{\partial^2}{\partial z\partial z^*}} \langle \zeta^{\H}_{z} |\hat H| \zeta^{\H}_{z} \rangle.
\end{equation}
 In order to obtain the function $H$ we have to invert Tuynman's relation (\ref{Tuynman-relation}):
  \begin{equation*} 
H-\frac{\partial^2 H}{\partial z \partial z^*} = h,
\end{equation*}
 which we can write formally as
   \begin{equation*} 
H = \biggl(\id-\frac{\partial^2 }{\partial z \partial z^*}\biggr)^{-1} h.
\end{equation*}
Inserting (\ref{hexp}) we obtain
  \begin{equation} 
H = \biggl(\id-\frac{\partial^2 }{\partial z \partial z^*}\biggr)^{-1} e^{-\frac{\partial^2}{\partial z\partial z^*}} \langle \zeta_{z} |\hat H| \zeta_{z} \rangle.
\end{equation}
We remark that $-\frac{\partial^2 f}{\partial z \partial z^*} =-\frac{\hbar}{2}\biggl( \frac{1}{\beta_0}\frac{\partial^2 f}{\partial q^2 } +{\beta_0}\frac{\partial^2 f}{\partial p^2 }   \biggr)$ is a positive operator  ($\sim -\Delta$ in adapted coordinates). Thus 
$ (\id-\frac{\partial^2 }{\partial z \partial z^*})^{-1}$ is well defined and bounded  in a suitably defined function space \cite{Tuynman_JMP87}. However, 
$ e^{-\frac{\partial^2}{\partial z\partial z^*}}$ is an unbounded operator, which is the origin of the regularity difficulties of the contravariant symbol. It will be regular if the contravariant symbol is in the domain of the Laplacian.

 \section{Conclusions}

 In summary, the formalism that  we have described allows one to construct models describing the interaction between classical and quantum systems in a well-defined Hilbert space framework.
 The geometric quantization of a classical system consists of selecting a subspace of the classical Hilbert space of functions on phase space.  The quantization of the observables is defined by projecting the classical observables on this subspace. The quantization of the dynamics involves first the addition of a dynamical and a geometrical phase to the classical dynamics and then projecting the generator of the dynamics on the subspace. The dequantization  of a quantum model  consists of the inverse procedure: given a Hamiltonian,  an algebra of observables represented in a Hilbert space, 
 and a set of coherent states, 
 one can construct an associated phase space manifold and the classical Hilbert space of square-integrable functions, with a suitable subspace that gives back the original quantum model when the geometric quantization is performed.

In the definition of the quantum models by Berezin-Toeplitz-geometrical quantization, Planck's constant $\hbar$ appears  in two places, that can be considered conceptually independent: The first one  is in the phase factor (\ref{koopman-phase}) of the pre-quantum Koopman-Schr\"odinger wave function. The second one  is in the selection of the polarization subspace, which depends crucially on the value of the constant $\hbarb$.  Although in principle the two constants  could be taken with two different independent values (to be determined by comparison with experiments), they are taken to be equal to a single constant $\hbar$. 
\\ \\

{ \bf Aknowledgements:}  We acknowlege support of the Marie Curie ITN Network FASTQUAST.
We thank M. Lachi\`eze-Rey, F. Faure and S. de Bi\`evre for very helpful discussions.

\end{document}